\documentclass[aps,prd,preprint,superscriptaddress,showpacs]{revtex4-1}

\usepackage{amsmath, amsfonts, amsthm, amssymb, mathrsfs, enumerate, graphicx}

\newcommand{\R}{{\mathbb R}}

\begin{document}


\title{A rotating universe outside a Schwarzschild black hole where spacetime itself non-uniformly rotates}


\author{Vee-Liem Saw}
\email[]{VeeLiem@ntu.edu.sg}
\affiliation{Division of Physics and Applied Physics, School of Physical and
Mathematical Sciences, Nanyang Technological University, 21
Nanyang Link, Singapore 637371}


\date{\today}

\begin{abstract}
We study a non-uniformly rotating universe outside a Schwarzschild black hole by generating a time-dependent manifold of revolution around a straight line. In this simple model where layers of spherical shells of the universe non-uniformly rotate, the Einstein field equations require this phenomenon to be caused by a static mass-energy distribution with time-dependent $T^{\phi\phi}$ (quadratic with time) and $T^{r\phi}=T^{\phi r}$ (linear with time). This indicates that a time-dependent stress along a certain direction results in a spacetime shift in that direction. For this model however, such material violates the null energy condition. Incidentally, the various coordinate systems describing the Schwarzschild solution can be viewed as arising from the freedom in parametrising the straight line and the radial function in the general method of constructing spacetime by generating manifolds of revolution around a given curve.
\end{abstract}


\maketitle


\section{Introduction}

The general method of constructing spacetime by generating manifolds of revolution around a given curve was recently formulated to study curved traversable wormholes \cite{Vee2013,Vee2012} \footnote{The original motivation that led to this formulation came from trying to helicalise a given curve \cite{Vee2013b}, i.e. replacing the given curve by another curve which winds around it.}. This follows from the ideas of Morris and Thorne \cite{Kip} that in finding solutions to general relativity, a spacetime geometry is first constructed and the Einstein field equations are subsequently used to determine the required materials to support it. Following the publication of Morris and Thorne's work, there has been intense research in the area of traversable wormholes, with a major result being the necessity of exotic matter to be present \cite{Vis2}. A useful application of the general method in \cite{Vee2013,Vee2012} is the construction of curved traversable wormholes which does not assume spherical symmetry. This led to the finding that by carefully engineering the shape and curvature of curved wormholes, it is possible for such wormholes to admit safe geodesics through them, i.e. freely-falling trajectories which are locally supported by ordinary matter, thereby avoiding the need for travellers to get into direct contact with exotic matter.

Here is how (3+1)-d spacetimes are constructed using that method. Given a smooth curve $\psi(v)$ embedded in a 4-d Euclidean space, a 3-manifold of revolution is:
\begin{eqnarray}\label{genmanifold}
\vec{\sigma}(u,v,w)=\psi(v)+Z(v)\cos{u}\ \vec{n}_1(v)+Z(v)\sin{u}\cos{w}\ \vec{n}_2(v)+Z(v)\sin{u}\sin{w}\ \vec{n}_3(v),
\end{eqnarray}
where $Z(v)$ is the radial function and $\vec{n}_1(v),\vec{n}_2(v),\vec{n}_3(v)$ are three orthonormal vectors. The metric of this 3-manifold can be calculated, and then extended to a (3+1)-d spacetime metric. This method can also be used to build dynamical spacetimes, as \cite{Vee2013} explicitly illustrated how an inflating wormhole can be constructed by letting the given curve $\psi(v)$ and the radial function $Z(v)$ depend on time.

In this paper, we would like to explore how the general method can be used to construct a rotating spacetime, and identify the essential attributes of the matter which produce this phenomenon. To investigate this in a simple model, extending a static traversable wormhole to a rotating one would not be ideal since it requires matter (exotic in some region) to already be present. It is instead advantageous to extend from an originally vacuum spacetime, since the matter that would be present is solely responsible for the rotational effects of the spacetime. Apart from the trivially flat Minkowski geometry, a Schwarzschild geometry is also vacuum (excluding the black hole, of course). We would hereby construct a rotating universe outside (the event horizon of) a black hole, where the spacetime itself rotates, and obtain the Minkowski version as the special case of zero black hole mass. To take full advantage of the spherical symmetry of the Schwarzschild geometry, the universe is prescribed to rotate in layers of rigid spherical shells. However, these shells of various radii need not be rotating with the same angular velocity (see Fig. 1), so the universe as a whole would not be rigidly rotating. This would be interesting, as we can compare such a non-uniformly rotating universe to an expanding universe which is isotropic and homogeneous (the FLRW solution), expanding in all directions whilst carrying the matter (or galaxies) along with it \cite{Gra}. Note also that we are going to assume that the mass-energy is static, unlike the van Stockum \cite{Stockum}, Tipler \cite{Tip} or Kerr \cite{VisserKerr} solutions which describe rotating matter. This helps to simplify the metric, since those examples necessarily contain a non-zero $g_{t\phi}$ cross-term which would lead to significantly arduous calculations and possibly obfuscate the interpretations of the matter properties, if incorporated into our model \footnote{As seen in the next section, the spatial 3-manifold representing a non-uniformly rotating universe would itself contain a $g_{v\phi}$ spatial cross-term.}.

\begin{figure}
\centering
\includegraphics[width=10cm]{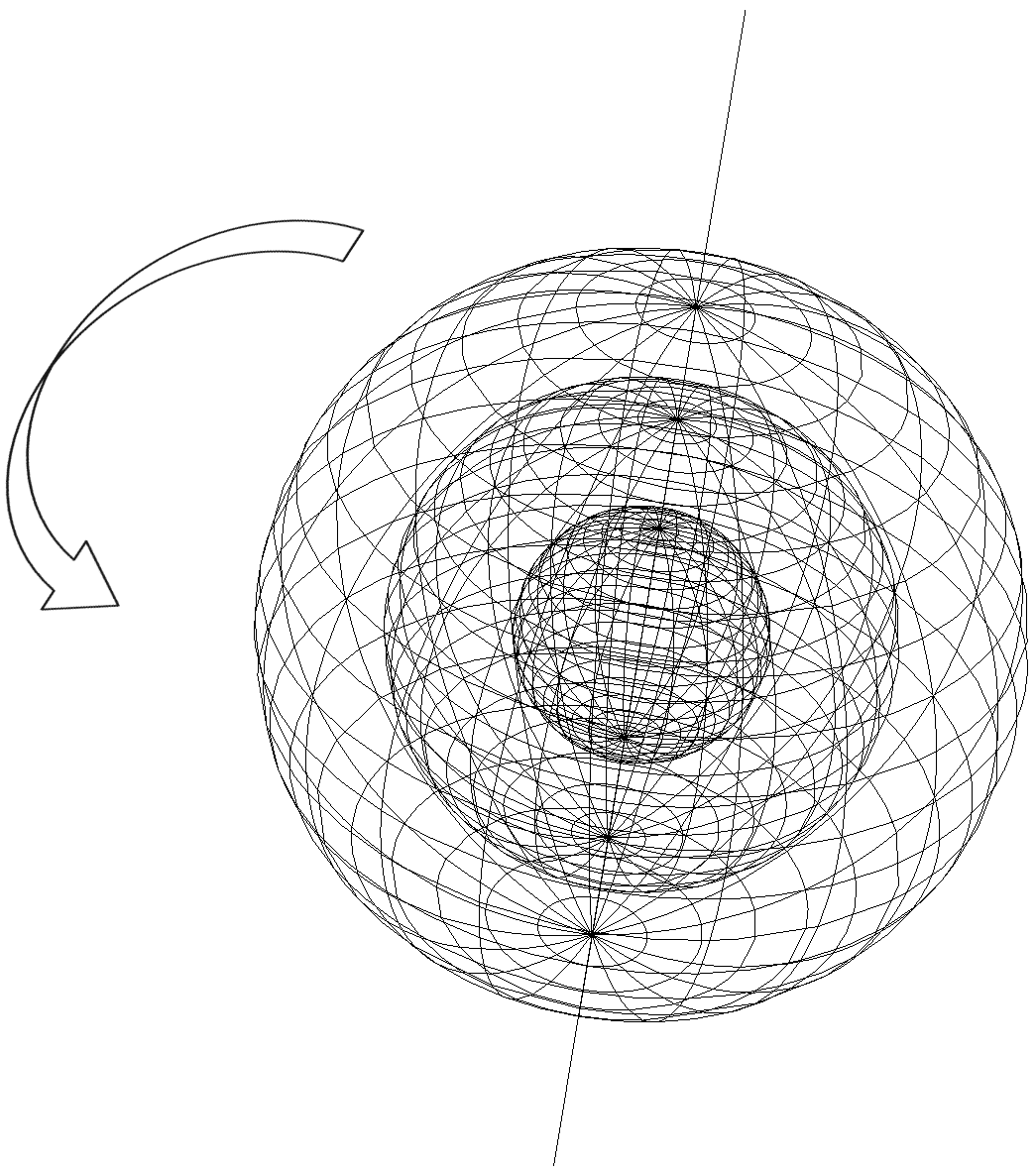}
\caption{Each spherical shell rigidly rotates, but spherical shells of different radii may be rotating with different angular velocities.}
\label{fig1}
\end{figure}

In the next section, we construct the metric that describes a non-uniformly rotating spacetime around a Schwarzschild black hole by generating a 3-manifold of revolution around a straight line. In section 3, we decide on the choice of parametrisation for the straight line and the radial function, showing how the freedom in parametrisation leads to various coordinate systems that describe the Schwarzschild metric. Section 4 is devoted to the physical properties of the mass-energy fluid that give rise to such a rotating spacetime, with a discussion section following after. Section 6 concludes this paper. We shall be working in units where $G=c=1$.

\section{The metric construction}

Consider the following time-dependent 3-manifold of revolution around a straight line embedded into a 4-d Euclidean space:
\begin{eqnarray}\label{rotmanifold}
\vec{\sigma}(t,v,\theta,\phi)=
\left(
\begin{array}{c}
r(v)\cos{\theta}\\
r(v)\sin{\theta}\cos{(\phi+\chi(v)\omega t)}\\
r(v)\sin{\theta}\sin{(\phi+\chi(v)\omega t)}\\
z(v)
\end{array}
\right),
\end{eqnarray}
where $t$ is the time coordinate as measured by a faraway observer, $r$, $\theta$ and $\phi$ being the usual spherical coordinates for 3-d Euclidean space, and $z$ is the fourth spatial coordinate \footnote{A useful lower-dimensional analogue for visualisation is to think of embedding a 2-d surface into ordinary 3-d Euclidean space using cylindrical coordinates $r$, $\phi$ and $z$ so that $r$ and $\phi$ are the usual plane polar coordinates with $z$ being the third spatial coordinate. The general framework for generating surfaces of revolution by adding circles or 1-spheres to a given curve is described in \cite{Vee2012}.}. The symbol $\omega$ is a constant, with $v$ parametrising the radial coordinate $r(v)$ and the fourth spatial coordinate $z(v)$, so it can be thought of that one is a function of the other, viz. $z(r)$ or $r(z)$. This effectively determines the shape of the manifold of revolution around the straight line.

Disregarding the $\chi(v)\omega t$ term, this 3-manifold of revolution obtained by adding spheres of radii $r(v)$ along the straight line $\vec{L}(v)=(0,0,0,z(v))$ is a special case of the general method used in \cite{Vee2013} to construct static curved traversable wormholes \footnote{Note that in \cite{Vee2013}, the coordinates $u$ and $w$ were used in place of $\theta$ and $\phi$, with the radial function denoted as $Z(v)$ (see Eq. (\ref{genmanifold})). The reason such general symbols were used was because they were not the usual spherical coordinates as the given curve (which was not necessarily a straight line since curved wormholes were to be constructed) would naturally induce a more general coordinate system. Here, the focus is on manifolds of revolution around a straight line (in fact rotating about that line) so we are indeed working with the three usual spherical coordinates $r,\theta,\phi$ plus the fourth spatial coordinate $z$.}. The time-dependence built in here by replacing $\phi\rightarrow \phi+\chi(v)\omega t$ from the static version in Eq. (\ref{genmanifold}) to yield Eq. (\ref{rotmanifold}) represents the fact that a spherical shell of radius $r(v)$ is rotating about the fourth coordinate axis $z$ (or the line $\vec{L}$) with constant angular velocity $\chi(v)\omega$. This can be seen by choosing any particular point on the 3-manifold i.e. fixing some values of $v,\theta,\phi$, and noting that as $t$ evolves this point would be rotated by an angle of $\chi(v)\omega t$ about $\vec{L}$. The $v$-dependence on $\chi$ implies that spheres of different radii $r(v)$ which are added at different points on $\vec{L}(v)$ may in general be rotating with different angular velocities, although we would require that this variation is smooth with $v$.

For physical interest that we would like to consider here in this study, the following conditions are imposed:
\begin{enumerate}[(i)]
\item
$\chi\rightarrow0$ as $v\rightarrow\infty$ so that when extended to a (3+1)-d spacetime manifold, a faraway observer would be sitting on an asymptotically flat manifold which is non-rotating. Physical observations shall be discussed with respect to this inertial frame.
\item
$\chi(v_0)$ is normalised to 1 at some reference point where the parameter is $v_0$. This reference spherical shell of radius $r(v_0)$ would then be rotating with angular velocity $\omega$.
\item
$\chi(v)$ monotonically decreases as $v$ goes from $v_0$ to $\infty$.
\end{enumerate}
The first two are boundary conditions that prohibit $\chi$ from being a constant (since it has to be 1 at $v_0$ and 0 at $\infty$), so the universe is rotating non-uniformly in contrast to being in a rigid rotation. This means that there is no observer that would see the universe as being globally static. Any observer that may be locally static would necessarily see at least one other spherical shell rotating. The third condition would demand that the bulk of the mass-energy be concentrated near the axis of rotation, diminishing away from it. Some kind of mass-energy distribution in the universe whose properties are to be found through the Einstein field equations would cause the universe itself to behave like a swirling fluid with maximum rate of swirling at the centre, and dissipating with essentially little or no swirling towards the outer edge. Note that this is not the same as the mass-energy being the said fluid that is rotating, rather it is the universe itself that is non-uniformly rotating \footnote{Like an expanding universe, the spacetime itself is expanding as opposed to the galaxies themselves moving away from each other.}.

The spatial metric $ds^2_{\textrm{space}}$ for the 3-manifold given by Eq. (\ref{rotmanifold}) can be computed as follows: the components are $g_{ij}=\vec{\sigma}_i\cdot\vec{\sigma}_j$, where $i,j\in\{v,\theta,\phi\}$, and $\vec{\sigma}_i$ denotes partial derivative with respect to $i$. This gives
\begin{eqnarray}
ds^2_{\textrm{space}}&=&(z'^2+r'^2+\chi'^2\omega^2t^2r^2\sin^2{\theta})dv^2+r^2d\theta^2+r^2\sin^2{\theta}\ d\phi^2\nonumber\\
&\ &+2\chi'\omega tr^2\sin^2{\theta}\ dvd\phi,
\end{eqnarray}
where explicit dependence on $v$ for $z(v),r(v),\chi(v)$ are suppressed for conciseness.
This spatial metric can also be written as,
\begin{eqnarray}
ds^2_{\textrm{space}}&=&(z'^2+r'^2)dv^2+r^2d\theta^2+r^2\sin^2{\theta}(\chi'\omega t\ dv+d\phi)^2,
\end{eqnarray}
indicating how the time-dependence term leads to the $dvd\phi$ cross-term, viz. $\phi\rightarrow\phi+\chi\omega t$ for the static to rotating manifold's parametric equations corresponds to $d\phi\rightarrow d(\phi+\chi\omega t)=d\phi+\chi'\omega t\ dv$ for their metrics. It is clear that a constant $\chi$ just gives the usual spherically symmetric metric (see section 7.2 in \cite{Vee2012}). Our boundary conditions however, forbid this for the non-uniform rotation case.

We can extend this to a spacetime metric of the form:
\begin{eqnarray}
ds^2&=&g_{tt}dt^2+(z'^2+r'^2)dv^2+r^2d\theta^2+r^2\sin^2{\theta}(\chi'\omega t\ dv+d\phi)^2,
\end{eqnarray}
where $g_{tt}<0$. Since we are chiefly concerned with the geometry outside a Schwarzschild black hole, $g_{tt}$ is a function of only $v$. We would not bother with the $g_{tj}$ ($j$ being any of the spatial coordinates $v,\theta,\phi$) cross-terms, taking them to be zero and impose the mass-energy to be static. This would greatly reduce the algebraic technicalities in calculating the Einstein tensor especially as the spatial metric itself already contains the $g_{v\phi}$ cross-term.

\section{Parametrisation of $z(v)$ and $r(v)$}

We should decide on the choice of parametrisation of $z(v)$ and $r(v)$, before proceeding with further computations. For the sake of discussion, let us consider the static spacetime so that $\omega=0$:
\begin{eqnarray}\label{staticmet}
ds^2&=&g_{tt}dt^2+(z'^2+r'^2)dv^2+r^2d\theta^2+r^2\sin^2{\theta}\ d\phi^2.
\end{eqnarray}
Amongst many possible parametrisations, two simple ones are linearly parametrising $r(v)=v$, or to linearly parametrise $z(v)=v$. The former is to treat the actual radial coordinate $r$ as the parameter itself, so that $z$ becomes a function of $r$ and the metric in Eq. (\ref{staticmet}) becomes
\begin{eqnarray}
ds^2&=&g_{tt}(r)dt^2+(z'(r)^2+1)dr^2+r^2d\theta^2+r^2\sin^2{\theta}\ d\phi^2.
\end{eqnarray}
If one goes on to calculate the Einstein tensor and solve the vacuum field equations, one would find that $z(r)=2\sqrt{R_s(r-R_s)}$ and $g_{tt}(r)=-(1-R_s/r)$, where $R_s$ is the Schwarzschild radius. This is the usual static spherically symmetric vacuum solution expressed in Schwarzschild coordinates and $z(r)=2\sqrt{R_s(r-R_s)}$ is recognised as Flamm's paraboloid.

If the latter parametrisation is used instead (which can also be thought of as linearly parametrising the line $\vec{L}$), then the metric in Eq. (\ref{staticmet}) becomes
\begin{eqnarray}
ds^2&=&g_{tt}(z)dt^2+(1+r'(z)^2)dz^2+r(z)^2d\theta^2+r(z)^2\sin^2{\theta}\ d\phi^2.
\end{eqnarray}
Solving the vacuum field equations gives $r(z)=z^2/4R_s+R_s$ and $g_{tt}(z)=-z^2/(z^2+4R_s^2)$. This is actually equivalent to the Einstein-Rosen coordinates if one rescales $z=2u\sqrt{R_s}$ \footnote{Matt Visser provides an excellent description on these various coordinate systems that cover only certain regions of the Schwarzschild geometry, which played a part in leading Einstein and Rosen to deduce the Einstein-Rosen bridge from their coordinates in \cite{EinRos}. The Einstein-Rosen coordinates are $ds^2=-u^2/(u^2+R_s)dt^2+4(u^2+R_s)du^2+(u^2+R_s)^2(d\theta^2+\sin^2{\theta}\ d\phi^2)$. See for instance chapter five in \cite{Vis2}.}. It is hereby obvious that they would have naturally interpreted from such coordinates that this represents a wormhole, since spheres of radii $r(z)=z^2/4R_s+R_s$ are added to the line $\vec{L}(z)=(0,0,0,z)$. With the radial function $r(z)>0$ for all $z\in\R$ and having a minimum value of $R_s$ at $z=0$, the geometrical picture of the spatial 3-manifold is a ``3-d straight tube'' with minimum radius $R_s$ at $z=0$ that grows into two asymptotically flat ends.

Those two parametrisations were rather effortless, i.e. setting either $r(v)=v$ or $z(v)=v$. As a third and perhaps not so straightforward example, isotropic coordinates can be obtained by the parametrisation $r(v)=v(1+R_s/4v)^2$ and $z(v)=(4v-R_s)\sqrt{R_s/4v}$ \footnote{The logical flow to arrive at this choice of parametrisation would be as follows: In isotropic coordinates, a spacetime metric takes the general form $ds^2=g_{tt}(v)dt^2+A(v)(dv^2+v^2d\theta^2+v^2\sin^2{\theta}\ d\phi^2)$. By comparing with our generic form in Eq. (\ref{staticmet}), we see that $z$ and $r$ has to satisfy $z'^2=r^2/v^2-r'^2$. By eliminating $z'^2$ in place of $r$ and $r'$, the vacuum field equation $G_{tt}=0$ reduces to $2v^2rr''=(r-vr')^2$, where $r(v)=v(1+R_s/4v)^2$ is a solution. It then follows that $z(v)=(4v-R_s)\sqrt{R_s/4v}$. Finally $g_{tt}(v)$ can be solved from other components of the vacuum field equations to give $g_{tt}(v)=-[(1-R_s/4v)/(1+R_s/4v)]^2$. The metric is hence $ds^2=-[(1-R_s/4v)/(1+R_s/4v)]^2dt^2+(1+R_s/4v)^4(dv^2+v^2d\theta^2+v^2\sin^2{\theta}\ d\phi^2)$, as anticipated. (See chapter 2.3.9 in \cite{Vis2}, for instance. Look out for the typo in Eq. (2.42) where $(1+r_s/4\rho)^2$ should be $(1+r_s/4\rho)^4$.)}. Fig. 2 shows how the same shape function is described by the three different parametrisations of $z(v)$ and $r(v)$ that are discussed here. It is thus intriguing that the various coordinate systems for the Schwarzschild geometry can be seen as arising from the freedom in parametrising $z(v)$ and $r(v)$ in the general method of constructing spacetime by generating manifolds of revolution around a given curve (in this case a straight line), as formulated in \cite{Vee2012,Vee2013} \footnote{In \cite{Vee2012,Vee2013} where static curved traversable wormholes were constructed, the parametrisations were essentially chosen from the beginning when the curve $\vec{\psi}(v)$ and the radial function $Z(v)$ were defined.}.

In our subsequent analysis of the rotating universe outside a Schwarzschild black hole, we shall adopt the parametrisation $r(v)=v$ which is the Schwarzschild coordinates.

\begin{figure}
\centering
\includegraphics[width=16cm]{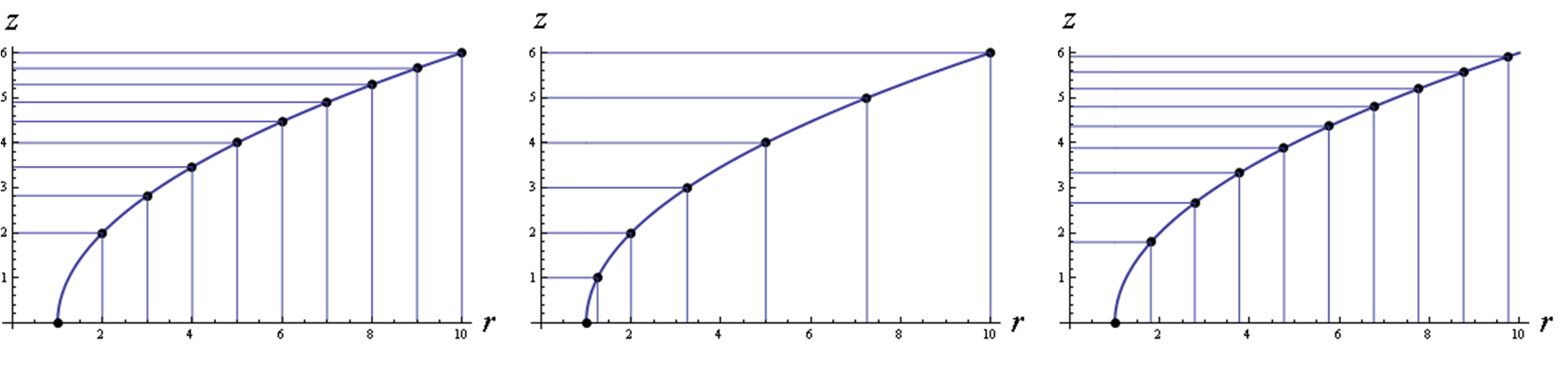}
\caption{The same curve (with $R_s=1$) as described by (from left) Schwarzschild coordinates ($r$ is linearly parametrised), Einstein-Rosen coordinates ($z$ is linearly parametrised), and isotropic coordinates. Each dot on the curve represents an increment of $v$ by 1, beginning from $(r,z)=(R_s,0)$.}
\label{fig2}
\end{figure}

\section{Physical properties of a non-uniformly rotating universe outside a Schwarzschild black hole}

Let us return to describing a non-uniformly rotating universe outside a Schwarzschild black hole with Schwarzschild radius $R_s$. In Schwarzschild coordinates, the spacetime metric would be
\begin{eqnarray}\label{Schmet1}
ds^2&=&-\left(1-\frac{R_s}{r}\right)dt^2+\frac{1}{1-R_s/r}dr^2+r^2d\theta^2+r^2\sin^2{\theta}(\chi'(r)\omega t\ dr+d\phi)^2\\\label{Schmet2}
&=&-\left(1-\frac{R_s}{r}\right)dt^2+\left(\frac{1}{1-R_s/r}+\chi'(r)^2\omega^2t^2r^2\sin^2{\theta}\right)dr^2+r^2d\theta^2+r^2\sin^2{\theta}\ d\phi^2\nonumber\\&\ &+2\chi'(r)\omega tr^2\sin^2{\theta}\ drd\phi,
\end{eqnarray}
where $r\geq R_s$ is the region of interest. The Einstein tensor $G^{\mu\nu}=R^{\mu\nu}-Rg^{\mu\nu}/2$ can be calculated, with the following non-zero terms \footnote{It is obvious that $\omega=0$ gives $G^{\mu\nu}=0$, so that a non-rotating Schwarzschild geometry is Ricci flat.}:
\begin{eqnarray}
G^{tt}&=&-\frac{\chi'^2\omega^2r^3\sin^2{\theta}}{4(r-R_s)}\\
G^{rr}&=&\frac{1}{4}\chi'^2\omega^2r(r-R_s)\sin^2{\theta}\\
G^{\theta\theta}&=&-\frac{1}{4}\chi'^2\omega^2\sin^2{\theta}\\
G^{\phi\phi}&=&\frac{1}{4}\chi'^2\omega^2\left(\chi'^2\omega^2t^2r(r-R_s)\sin^2{\theta}-3\right)\\
G^{r\phi}&=&G^{\phi r}=-\frac{1}{4}\chi'^3\omega^3tr(r-R_s)\sin^2{\theta}\\
G^{t\phi}&=&G^{\phi t}=-\frac{\omega}{2r}(4\chi'+r\chi'')
\end{eqnarray}
A particularly interesting quick observation is that $G^{t\phi}=G^{\phi t}$ can be made to be identically zero if $\chi$ satisfies $4\chi'+r\chi''=0$. A solution to this is $\chi(r)=P/r^3+Q$ where $P,Q$ are arbitrary constants. The two boundary conditions for $\chi$ (see section 2) would be met if $P=R_s^3$ and $Q=0$, with the reference spherical shell being the horizon of the black hole $\chi(R_s)=1$ since the region observable by a faraway observer is $r\geq R_s$. The third condition is also met by $\chi(r)=(R_s/r)^3$. This leaves the non-zero components of $G^{\mu\nu}$ as
\begin{eqnarray}
G^{tt}&=&-\frac{9R_s^6\omega^2\sin^2{\theta}}{4r^5(r-R_s)}\\
G^{rr}&=&\frac{9R_s^6\omega^2(r-R_s)\sin^2{\theta}}{4r^7}\\
G^{\theta\theta}&=&-\frac{9R_s^6\omega^2\sin^2{\theta}}{4r^8}\\
G^{\phi\phi}&=&\frac{27}{4}R_s^6\omega^2\left(\frac{3R_s^6\omega^2t^2(r-R_s)\sin^2{\theta}-r^7}{r^{15}}\right)\\
G^{r\phi}&=&G^{\phi r}=\frac{27R_s^9\omega^3t(r-R_s)\sin^2{\theta}}{4r^{11}}.
\end{eqnarray}

The physics of the mass-energy that would result in such a non-uniformly rotating universe is given by the Einstein field equations $G^{\mu\nu}=8\pi T^{\mu\nu}$. With $\chi=(R_s/r)^3$, the $T^{t\phi}=T^{\phi t}$ terms are zero, so the simplest kind of mass-energy does not involve any heat transfer. In full, the non-zero components of $T^{\mu\nu}$ are:
\begin{eqnarray}\label{genTbegin}
T^{tt}&=&-\frac{9R_s^6\omega^2\sin^2{\theta}}{32\pi r^5(r-R_s)}\leq0\\
T^{rr}&=&\frac{9R_s^6\omega^2(r-R_s)\sin^2{\theta}}{32\pi r^7}\geq0\\
T^{\theta\theta}&=&-\frac{9R_s^6\omega^2\sin^2{\theta}}{32\pi r^8}\leq0\\
T^{\phi\phi}&=&\frac{27}{32\pi}R_s^6\omega^2\left(\frac{3R_s^6\omega^2t^2(r-R_s)\sin^2{\theta}-r^7}{r^{15}}\right)\\\label{genTend}
T^{r\phi}&=&T^{\phi r}=\frac{27R_s^9\omega^3t(r-R_s)\sin^2{\theta}}{32\pi r^{11}}\geq0\textrm{\ for\ }t\geq0.
\end{eqnarray}

A reassuring fact that can be inferred regarding the nature of the mass-energy density $T^{tt}$ is its time-independence. Note the contrast when compared to an expanding universe which dilutes the static material as it carries it along in the expansion. Here, there is no increase in volume since spherical shells of the universe are rotating but not expanding. The material density is naturally expected to be constant with time, lest mass-energy conservation be violated to produce the rotation.

As seen from the frame of a faraway observer, the mass-energy density $T^{tt}$ is negative and gets enormously large towards the horizon as $\displaystyle \lim_{r^+\rightarrow R_s}T^{tt}=-\infty$, though the $T^{rr}$ and $T^{r\phi}=T^{\phi r}$ stresses vanish at $r=R_s$. There are two (independent) components of the stress-energy tensor which are time-dependent, viz. $T^{\phi\phi}$ and $T^{r\phi}=T^{\phi r}$, where the former depends quadratically with time and linearly for the latter. The other two stresses $T^{rr}$ and $T^{\theta\theta}$ are time-independent. The field equations therefore reveal the astonishing effect of a mass-energy fluid whose stresses possess this directional dependence on time: Such static fluid, with time-dependent properties of $T^{\phi\phi}$ and $T^{r\phi}=T^{\phi r}$ would remarkably cause the universe to rotate, carrying the fluid along with it.

\begin{figure}
\centering
\includegraphics[width=8cm]{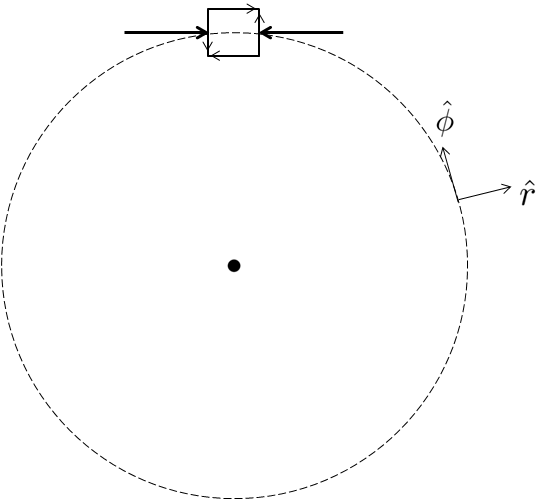}
\caption{The time-dependent stresses on a fluid element $T^{\phi\phi}$ (quadratic with time) and $T^{r\phi}=T^{\phi r}$ (linear with time) in the $\theta=\pi/2$ plane. The frame of a faraway observer would see the dotted circle trajectory of the static fluid element being carried along by the universe which rotates about a black hole. The time-independent stresses along the radial and $\theta$-directions are not shown.}
\label{fig3}
\end{figure}

Although a free particle cannot remain at rest outside a black hole (since the gravity of the black hole would attract the particle towards it), it is not difficult to imagine some kind of cosmological event where perhaps a red giant exploded and began collapsing into a black hole. During the explosion, the red giant would expel material outwards, eventually leading to a transient equilibrium state where the material is static outside the resulting black hole's event horizon for a period of time. The sign of $T^{\phi\phi}$ is negative for small $t$, indicating that the material is initially under tension along the $\phi$-direction. Over time, the sign of $T^{\phi\phi}$ changes to positive such that the material would be under increasing pressure. An example of a material whose pressure increases would be a nuclear process, whereby mass is converted into thermal energy which builds up the pressure over time, though in this situation the pressure increase only takes place along the $\phi$-direction and the $r\phi$-shears. A realistic scenerio of this non-uniformly rotating phenomenon would only be temporary, since the increase cannot go unbounded forever. Fig. 3 depicts the time-dependent stresses on a fluid element with increasing pressure along the $\phi$-direction and $r\phi$-shear forces.

Finally, consider the covariant null vector $k_\mu=((\sqrt{-g^{tt}})^{-1},0,0,(\sqrt{g^{\phi\phi}})^{-1})$. For $\theta\neq\pi/2$,
\begin{eqnarray}
T^{\mu\nu}k_\mu k_\nu=T^{tt}(k_t)^2+T^{\phi\phi}(k_\phi)^2=-\frac{9R_s^6\omega^2r\sin^2{\theta}}{8\pi[r^7+9R_s^6\omega^2(r-R_s)t^2\sin^2{\theta}]}<0,
\end{eqnarray}
implying that the null energy condition is violated \footnote{The null energy condition asserts that for any null vector, $T^{\mu\nu}k_\mu k_\nu\geq0$. This can be found in page 115 of \cite{Vis2}.}. This kind of mass-energy is therefore exotic in order to produce this non-uniformly rotating universe. The negativity of $T^{\mu\nu}k_\mu k_\nu$ for this null vector however, decreases with time.

\subsection{Minkowski spacetime ($R_s=0$)}

A non-uniformly rotating Minkowski spacetime can be thought of as the special case when the mass of the black hole is zero, or equivalently $R_s=0$. We cannot however, directly substitute $R_s=0$ into the stress-energy tensor because we chose $P=R_s^3$ to satisfy our two boundary conditions for $\chi$. To obtain the correct $T^{\mu\nu}$, we let $\chi=(R_0/r)^3$ with $R_0$ being a positive constant and only consider the region where $r\geq R_0$. The non-zero components of $T^{\mu\nu}$ corresponding to Eqs. (\ref{genTbegin}-\ref{genTend}) are
\begin{eqnarray}
T^{tt}&=&-\frac{9R_0^6\omega^2\sin^2{\theta}}{32\pi r^6}\leq0\\
T^{rr}&=&\frac{9R_0^6\omega^2\sin^2{\theta}}{32\pi r^6}\geq0\\
T^{\theta\theta}&=&-\frac{9R_0^6\omega^2\sin^2{\theta}}{32\pi r^8}\leq0\\
T^{\phi\phi}&=&\frac{27}{32\pi}R_0^6\omega^2\left(\frac{3R_0^6\omega^2t^2\sin^2{\theta}-r^6}{r^{14}}\right)\\
T^{r\phi}&=&T^{\phi r}=\frac{27R_0^9\omega^3t\sin^2{\theta}}{32\pi r^{10}}\geq0\textrm{\ for\ }t\geq0.
\end{eqnarray}

As in the universe outside a Schwarzschild black hole, the null energy condition is violated since a null vector $k_\mu=((\sqrt{-g^{tt}})^{-1},0,0,(\sqrt{g^{\phi\phi}})^{-1})$ gives (for $\theta\neq\pi/2$)
\begin{eqnarray}
T^{\mu\nu}k_\mu k_\nu=T^{tt}(k_t)^2+T^{\phi\phi}(k_\phi)^2=-\frac{9R_0^6\omega^2\sin^2{\theta}}{8\pi(r^6+9R_0^6\omega^2t^2\sin^2{\theta})}<0.
\end{eqnarray}

The results for the non-uniformly rotating Minkowski spacetime are therefore similar to that for Schwarzschild. Note however that unlike Schwarzschild where $r\geq R_s$ is the observable universe, here the entire spacetime should be observable. This ostensibly leads to a problem, because $r\geq R_0\neq0$ otherwise $\chi=(R_0/r)^3$ would then be identically zero. Nevertheless the form of $\chi=P/r^3+Q$ was a solution to $G^{t\phi}=G^{\phi t}=0$ such that $T^{t\phi}=T^{\phi t}=0$ which gives a stress-energy tensor that does not involve heat conduction. It is certainly possible to choose a different $\chi$ which is 1 at $r=0$ and monotonically decreases to 0 as $r\rightarrow\infty$, like $\chi=1/(1+r^3)$ but requires that $T^{t\phi}=T^{\phi t}\neq0$. This is thus a difference between a non-uniformly rotating Minkowski and a non-uniformly rotating Schwarzschild universe. Furthermore, the mass-energy for the Minkowski one is naturally static since there is no black hole to gravitationally attract it towards the centre \footnote{If $R_s=0$, then from Eqs. (\ref{Schmet1}) or (\ref{Schmet2}) the metric for a non-uniformly rotating Minkowski spacetime has $g_{tt}=-1$, which implies no tidal force.}.

\section{Further Discussion}

We began by constructing the geometry of a non-uniformly rotating universe around a Schwarzschild black hole in section 2, and subsequently showed that this is caused by a static mass-energy with time-dependent stress along the $\phi$-direction and $r\phi$-shear forces. The frame of a faraway observer sees that the mass-energy is negative, and violates the null energy condition. The exotic nature of the material may render this phenomenon as unphysical, notwithstanding the fact that there are known solutions in general relativity (like traversable wormholes \cite{Vis2}, and warp drive \cite{Alc}) which demand such physics. Whilst it is arguable that the Casimir effect \cite{Exo1,Exo2,Exo3} is a well-regarded example of exotic matter to support these kind of so-called unphysical solutions, it is certainly important to be critical and discrimate any artificially thought up spacetime with highly obscure and inordinate physical requirements.

In spite of the possibility of being classified into the undesirable category, the purpose of our study here is not to propose an arbitrary metric and just accept whatever stress-energy tensor that follows from the field equations. It is actually quite the contrary as we do not demand that there must exist a particular kind of mass-energy in nature to produce our desired spacetime. Instead, our motivation lies in figuring out the properties of such material and uncover their key characteristics. If such properties are deemed drastically preposterous, then it may perhaps be interpreted as an explanation to why we do not observe such rotational effects in our universe.

The Einstein field equations are notoriously complicated non-linear partial differential equations, and a repercussion is the difficulty to solve it exactly. They are nevertheless meant to be read both ways:
\begin{enumerate}[(i)]
\item
Given the physics, here is the resulting spacetime geometry.
\item
Given a spacetime geometry, here is the necessary physics.
\end{enumerate}
Even though our approach is based on the unorthodox direction of specifying the spacetime geometry to figure out the necessary physics, our results can be read from the more conventional direction to reveal the effect of a static mass-energy with time-dependent $\phi$-stress (quadratic with time) and $r\phi$-shear forces (linear with time): \emph{This causes the universe to rotate, i.e. shift along the $\phi$-direction.} Putting it in another way, the stress-energy time-dependence for a particular direction ($T^{\phi\phi}$, $T^{r\phi}=T^{\phi r}$) results in a spacetime translation along that direction ($\phi$). This may indicate that more complicated dynamical evolution of the universe can be decomposed into the respective directional time-dependence of the stress-energy tensor ($T^{rr}(t),T^{\theta\theta}(t)$, etc.). Our success in pinning down the precise conditions for this particular case owes to the fact that we constructed such a rotating spacetime first and then use the field equations to decipher the physics. We therefore already have an exact solution, for what may appear to be a strange (or highly fine-tuned) specification of the stress-energy tensor. Surely, one may begin instead with a physically constructed stress-energy tensor with the time-dependent properties of $T^{\phi\phi}$ and $T^{r\phi}=T^{\phi r}$. The major weakness in this usual approach is that unless the stress-energy is of a particularly nice form, it may be nearly impossible to analytically solve the field equations. An important lesson is thus had we remained obdurate and refused to be avant-garde with the field equations, we might not have discovered such properties that lead to a rotating universe.

Our results also show the difference between a non-uniformly rotating universe and an expanding universe as described by the FLRW solution. The latter assumes homogeneity and isotropicity, so it does not pick out any preferred spatial direction. The resulting Friedmann equations govern the change in the material density with time as it causes the universe to expand. Our non-uniformly rotating universe on the other hand has a fixed density and it is the increase in a directional stress with time that produces the rotation.

For future research, it would be interesting to investigate if the purported exotic nature of the material to produce the rotation is mandatory. One may attempt to adapt the key time-dependent features that we have found here to specify certain stress-energy tensors (perhaps with no or less severe violation of the null energy condition), and solve the field equations numerically to simulate the rotational evolution of the universe. Our model assumes a constant angular velocity for simplicity to glean the crucial physical insights. It would be desirable to study more exhaustively how the rotation may originate, and how it would end via numerical computations. Another possible extension would be to let the material orbit the black hole on a circular geodesic, although care is to be taken in distinguishing between the particle's own orbital motion with the rotation of the universe itself.

This simple model presented here is based on Einstein's theory of general relativity, without any quantum effects involved. Recent frontier research in quantum gravity has pushed the debate on what happens near the horizon of a black hole to unprecedented heights, following the black hole information paradox \cite{BHW,Haw1,Susskinda,Hooft,Susskindb} to firewalls near a black hole horizon \cite{nature,Firewall1}, with the latest update from Stephen Hawking suggesting that ADS-CFT supports the notion that black holes do not have event horizons \cite{Hawkinglatest}. It would be exciting to extend the formulation of a non-uniformly rotating universe around a black hole to the realm of such theories where quantum mechanics plays its part as well.

\section{Concluding remarks}

This study of a non-uniformly rotating universe around a Schwarzschild black hole is perhaps a paramount example of how the approach of constructing the spacetime metric based on its desired geometrical properties and then using the field equations to find out the physics of it has led to the discovery of a new kind of solution in general relativity. The key properties of the mass-energy for this spacetime have been carefully examined and discussed.

We have also illustrated how our general method of constructing spacetime by generating manifolds of revolution around a curve \cite{Vee2012,Vee2013} leads to the various coordinate systems for describing the Schwarzschild solution, when the given curve is a straight line. These different coordinate systems can be attributed to the freedom in parametrising the straight line and the radial function.

\begin{acknowledgments}
The author is grateful for the efforts by Meng Lee Leek from Nanyang Technological University in participating in useful discussions that aided towards the coherent formulation of this work, as well as reviewing the final draft of this manuscript.
\end{acknowledgments}

\bibliographystyle{spphys}       
\bibliography{Citation}

\end{document}